\documentstyle[aps,prl,twocolumn,psfig,floats]{revtex}     

\begin{document}
\preprint{\vbox{\hbox{{\tt hep-th/0210294}\\ LPT-ORSAY 02-95\\ October 2002}}}
\draft
\wideabs{
\title{GUT Precursors and Non-Trivial Fixed Points\\ in Higher-Dimensional Gauge Theories}
\author{Keith R. Dienes$^1$, Emilian Dudas$^{2,3}$, Tony Gherghetta$^4$}
\address{$^1$ Department of Physics, University of Arizona, Tucson, AZ  85721 USA}
\address{$^2$ LPT, B\^at.~210, Univ.\ Paris-Sud, F-91405, Orsay Cedex, France}
\address{$^3$ Centre de Physique Th\'eorique, Ecole Polytechnique, F-91128, Palaiseau Cedex, France}
\address{$^4$ School of Physics and Astronomy, University of Minnesota,
               Minneapolis, MN 55455 USA}
\date{October 29, 2002}
\maketitle
\begin{abstract}
      Within the context of traditional logarithmic grand unification at
      $M_{\rm GUT}\approx 10^{16}$~GeV, we show that it is nevertheless
      possible to observe certain GUT states such as $X$ and $Y$ gauge
      bosons at lower scales, perhaps even in the TeV range.  We refer
      to such states as ``GUT precursors.''  Such
      states offer an interesting alternative possibility for new
      physics at the TeV scale, even when the scale of gauge 
      coupling unification remains high,
      and suggest that it may be possible to probe
      GUT physics directly even within the context of high-scale
      gauge coupling unification.
      More generally, our results also suggest that it is possible to construct
      self-consistent ``hybrid'' models containing widely separated energy scales,
      and give rise to a Kaluza-Klein realization of non-trivial 
      fixed points in higher-dimensional gauge theories.
      We also discuss how such theories may be deconstructed at high energies.
\end{abstract}
\bigskip
\bigskip
          }

\newcommand{\newc}{\newcommand}
\newc{\gsim}{\lower.7ex\hbox{$\;\stackrel{\textstyle>}{\sim}\;$}}
\newc{\lsim}{\lower.7ex\hbox{$\;\stackrel{\textstyle<}{\sim}\;$}}

\def\beq{\begin{equation}}
\def\eeq{\end{equation}}
\def\beqn{\begin{eqnarray}}
\def\eeqn{\end{eqnarray}}
\def\calM{{\cal M}}
\def\Nbar{{\overline{N}}}
\def\half{{\textstyle{1\over 2}}}
\def\ie{{\it i.e.}\/}
\def\eg{{\it e.g.}\/}
\def\etc{{\it etc}.\/}


\def\inbar{\,\vrule height1.5ex width.4pt depth0pt}
\def\IR{\relax{\rm I\kern-.18em R}}
 \font\cmss=cmss10 \font\cmsss=cmss10 at 7pt
\def\IQ{\relax{\rm I\kern-.18em Q}}
\def\IZ{\relax\ifmmode\mathchoice
 {\hbox{\cmss Z\kern-.4em Z}}{\hbox{\cmss Z\kern-.4em Z}}
 {\lower.9pt\hbox{\cmsss Z\kern-.4em Z}}
 {\lower1.2pt\hbox{\cmsss Z\kern-.4em Z}}\else{\cmss Z\kern-.4em Z}\fi}

\input epsf


\section{Introduction}

One of the most important theoretical challenges in physics is to 
determine the nature of fundamental theories.
Such fundamental theories include theories of grand unification,
quantum gravity, and even strings, with each theory carrying its own intrinsic
energy scale.  

The traditional view of such theories stipulates that their intrinsic energy scales
are exceedingly high.  In such cases, experimental evidence
in favor of such theories is at best indirect.
More recently, however, it has been suggested~\cite{TeV1,TeV2,DDG,TeV4} that 
the presence of large extra dimensions might significantly lower the energy scales 
associated with such theories, perhaps all the way to the TeV range. 
In such cases, we might hope for direct
experimental tests of such theories.  

In this paper, we wish to propose a ``hybrid'' possibility.  
Specifically, we wish to consider a higher-dimensional scenario in which 
the fundamental theories of physics retain their traditional high characteristic 
energy scales, but in which it is nevertheless possible to obtain {\it direct}\/, 
low-energy evidence of their existence.  
As we shall see, this will be possible because of the emergence 
of a non-trivial fixed point which enables a large separation of scales to exist
within a single model.

For concreteness, we shall concentrate on the case of grand unification, and
consider a scenario in which the unification of gauge couplings retains its
traditional logarithmic behavior, with unification occurring near 
$M_{\rm GUT}\approx 10^{16}$~GeV.  This unification is therefore precisely 
as in the Minimal Supersymmetric Standard Model (MSSM).
However, we shall demonstrate that even within such a scenario, it is possible
that certain states associated with the emergence of a grand unified
theory (GUT) at this energy scale can actually be extremely light, perhaps even
in the TeV range.  
We shall refer to such states as ``GUT precursors.'' 
The appearance of such precursor states would then provide a direct, experimental 
window into high-scale, fundamental physics.

\section{GUT symmetry breaking, orbifolds, and ``GUT precursors''}

In theories which exhibit a unification of the Standard Model (SM) gauge couplings, 
it is natural to imagine the
emergence of a grand unified theory 
at the scale of unification.  
The gauge symmetry group associated with this GUT [\eg, $SU(5)$ or $SO(10)$] must then be large enough
to contain the SM gauge symmetry group as a subgroup.  
In each case, we then find that the corresponding GUT contains not only
the usual Standard Model particles, but also additional particles which
are directly associated with the GUT.  These particles necessarily include the so-called
$X$ and $Y$ gauge bosons associated with the enlarged GUT gauge symmetry,   
and may also include additional matter particles (such as colored Higgs triplets).

There are two basic methods by which GUT symmetries can be broken below the scale 
of unification.
The first method is intrinsically field-theoretic:  one imagines that a certain GUT
field obtains a non-vanishing expectation value $v\approx M_{\rm GUT}$ in such 
a way that the Standard Model fields remain light while the extra GUT
fields become heavy.  This is the standard Higgs mechanism for breaking
a GUT symmetry, and it is characterized by the fact that       
the masses of the extra GUT fields ---the true signatures of the existence of
the GUT--- are parametrically tied to the GUT scale.
  
The second method, by contrast, is essentially string-theoretic, and
involves truncating the full string Fock space 
in such a way that the large initial gauge symmetry is broken 
down to a smaller residual gauge symmetry.  This method has a long history in the 
string literature~\cite{Witten,strings}, and is the method
by which large string gauge symmetries such as $SO(32)$ or $E_8\times E_8$
are broken down to the Standard Model gauge group in various phenomenological
string models~\cite{Faraggi,models}. 
This method has often been referred to as ``GUT breaking by orbifolds,''
and has been discussed within the context of  
large extra dimensions in Ref.~\cite{DDG} as well as more
recently in Refs.~\cite{Kawamura,HN,JMR,otherHN}.

This method works as follows.
For simplicity, let us first imagine compactification 
of a single extra dimension on the circle $S^1$ defined by identifying
$y\approx y+2 \pi R$ 
(where $y$ is the coordinate along the compact extra dimension).  
Any field $\Phi(x^\mu,y)$ on the circle can then be Fourier-decomposed as
$\Phi=\Phi_+ + \Phi_-$ where
\beqn
     \Phi_+(x^\mu,y) &=& \sum_{n=0}^\infty \Phi_+^{(n)}(x^\mu) 
                 \cos\left( {ny\over R}\right)~,\nonumber\\
     \Phi_-(x^\mu,y) &=& \sum_{n=1}^\infty \Phi_-^{(n)}(x^\mu) 
                  \sin\left( {ny\over R}\right)~.
\label{KKdecomp}
\eeqn
Note that $\Phi_+$ is even under $y\to -y$, while $\Phi_-$ is odd;  moreover $\Phi_-$
lacks a zero mode.
The mass of the Kaluza-Klein mode $\Phi_\pm^{(n)}$ is $n/R$.
Given this Kaluza-Klein circle decomposition, it is then straightforward to  
compactify on the orbifold defined by $S^1/Z_2$ where the $Z_2$ action is $y\to -y$:
we simply retain only the even or odd modes in the above decomposition.
It is this truncation which reduces the Fock space.
For example, if $\Phi$ refers to a Standard Model field, we arrange
our orbifold so as to retain the even components $\Phi_+$, whereas if 
$\Phi$ refers to a GUT field which is not present in the Standard Model,
we retain the odd components $\Phi_-$.
In this way, the spectrum of zero-modes accessible to the low-energy observer at
energies $E\ll R^{-1}$ consists of only the Standard Model fields.  In other words,
the orbifold projection has broken the GUT at low energies.

For our purposes, however, the important feature of this method
of GUT symmetry breaking is the energy scale at which the first signatures
of the full GUT symmetry appear.  Unlike the Higgs breaking mechanism,
where masses of the GUT fields beyond the Standard Model 
are parametrically tied to $M_{\rm GUT}$,
in this case
the masses of the first Kaluza-Klein modes for these GUT particles 
are set by the inverse radius of the orbifold!
Thus, in cases for which $R^{-1} <M_{\rm GUT}$,
we actually begin to observe GUT particles (such as $X$ and 
$Y$ gauge bosons) {\it before}\/ we detect actual gauge coupling
unification.  In other words, these low-lying Kaluza-Klein modes 
of the GUT particles appear as ``GUT precursors''~\cite{DDG}, signalling the 
future emergence of a full gauge coupling unification 
at an even higher energy scale.

The obvious question, then, is to determine how light these
GUT precursors can be.  How far below $M_{\rm GUT}$, the
scale of gauge coupling unification, can these states sit?
 
Clearly, the answer to this question depends on the particular model
under discussion.  In the models of accelerated power-law unification in 
Ref.~\cite{DDG}, 
the ratio between the scale of
extra dimensions and the scale of accelerated gauge coupling unification 
is never significantly more than one order of magnitude:  
$M_{\rm GUT} R \lsim 20$.
Likewise, in the more recent models of Ref.~\cite{HN} for which
$M_{\rm GUT}$ takes a high value $M_{\rm GUT} > 10^{16}$~GeV, this
ratio is somewhat larger:  $M_{\rm GUT} R \approx 100$.
Thus, both classes of models predict the appearance of 
GUT precursors well in advance of actual gauge coupling unification.

Despite these facts, both classes of models predict the appearance
of GUT precursors which are not drastically separated 
from their corresponding fundamental scales of gauge coupling
unification.  In other words, neither class of models provides for an
extremely large energy range over which the effective theory
is higher-dimensional with non-unified gauge couplings.

At first glance, it might appear that we cannot separate these
scales too greatly because the GUT precursors themselves will
affect the running of the gauge couplings and thereby
alter the gauge coupling unification which is responsible
for setting the value $M_{\rm GUT}$.
Indeed, 
the cumulative effects of the Kaluza-Klein excitations  
of {\it Standard Model}\/ fields give rise to a power-law running
for the gauge couplings which tends to accelerate the scale of 
gauge coupling unification, thereby tying $M_{\rm GUT}$ closely to 
the precursor scale $R^{-1}$.

However, when the effects of the GUT precursors are also included, 
it then follows that the states at each excited Kaluza-Klein mass level
fall into complete GUT multiplets.
As originally noticed in Ref.~\cite{DDG}, this still
leads to gauge coupling unification.  However, as pointed out 
in Ref.~\cite{HN}, the presence of complete GUT multiplets at each
excited level implies that the power-law running is {\it universal}\/ for
each gauge coupling.  Thus, the {\it unification}\/ of the gauge couplings
continues to be logarithmic, occurring just as it does in four dimensions
in the absence
of Kaluza-Klein states.  Indeed, this feature is the hallmark 
of the models of Ref.~\cite{HN}, and is one of the reasons
why these models can apparently tolerate the larger value   
$M_{\rm GUT} R \approx 100$.

The chief danger inherent in such a unification, however, is that the
gauge couplings are each individually still experiencing power-law evolution.
Thus, even though the {\it unification}\/ of these gauge couplings is
completely logarithmic, the gauge couplings
themselves might flow towards strong coupling, thereby invalidating
the perturbative unification calculation.    
Indeed, this is the primary feature which ultimately limits the separation
between the scale at which the GUT precursors appear and the 
scale of gauge coupling unification.

\section{Power-Law Running and Perturbativity}

We shall now demonstrate that this restriction
does not arise for models in which all matter is restricted
to orbifold fixed points and in which only the Standard Model and
GUT gauge bosons
propagate in the bulk.  In such cases, we shall show that
it is possible to separate $R^{-1}$, the scale of the GUT 
precursors, {\it by an arbitrary amount}\/ from $M_{\rm GUT}$,
at least as far as gauge coupling unification and perturbativity
are concerned.  We shall defer our discussion of the general interpretation 
and phenomenological implications 
of these results to Sects.~IV and VII.

In theories with extra dimensions, the evolution of the gauge couplings takes the 
approximate form~\cite{DDG}
\beqn
       \alpha_i^{-1} (\Lambda)  &\approx &
       \alpha_i^{-1} (M_Z) 
      - {b_i\over 2\pi} \ln {\Lambda\over M_Z} +
     {\tilde b_i\over 2\pi} \ln {\Lambda R}\nonumber\\
         &&~~~~~~ -
     {\tilde b_i X_\delta \over 2\pi \delta} \left\lbrack
                (\Lambda R)^\delta - 1 \right\rbrack~.
\label{RGE}
\eeqn
The emergence of power-law behavior 
is expected in a higher-dimensional gauge theory~\cite{TV,DDG},
and can equivalently be viewed as logarithmic running 
with a beta-function coefficient that continually changes
as successive Kaluza-Klein thresholds are crossed.
In Eq.~(\ref{RGE}), $M_Z$ is our chosen low-energy reference scale;  $\Lambda$ is an arbitrary
high scale (ultimately associated with the cutoff of the higher-dimensional 
gauge theory);
$\delta$ is the number of compactified dimensions;
$R$ is their common radius of compactification;  
and the normalization factor $X_\delta$ is the 
compactification volume with all radii normalized to unity.
For example, for toroidal compactifications we have $X_\delta \equiv \pi^{\delta/2}/\Gamma(1+\delta/2)$
(the volume of a $\delta$-dimensional unit sphere).
Likewise, $b_i$ are the beta-function coefficients of the zero-mode fields (including
the contributions of those fields which do not feel extra dimensions altogether,
such as those restricted to branes and/or orbifold fixed points), while $\tilde b_i$
are the beta-function coefficients associated with the field content at 
each excited Kaluza-Klein level.  
Thus, it is the latter beta-function coefficients which govern the 
intrinsically higher-dimensional power-law evolution of the gauge couplings.
Note that in the case of compactification on an orbifold, 
the $\tilde b_i$-dependent contributions in Eq.~(2) are generally
reduced due to the orbifold projections at the excited Kaluza-Klein
levels;  with a $\IZ_2$ projection, for example, these terms must be divided by $2$.
Also note that if the zero-mode fields are those of the MSSM, then the sum of the
first two terms on the right side of Eq.~(\ref{RGE}) is equal 
to the usual MSSM value of the (inverse) unified gauge coupling
when $\Lambda=M_{\rm GUT}\approx 2\times 10^{16}$~GeV.

In cases where the $\tilde b_i$ are unequal, we see from Eq.~(\ref{RGE}) that
the power-law evolution of
the gauge couplings is different for each gauge coupling.  This implies
that the relative {\it differences}\/ between the gauge couplings 
also evolve with power-law behavior.
However, when the $\tilde b_i$ are all equal, 
we see from Eq.~(\ref{RGE}) that
this power-law behavior is universal for all gauge couplings.  
The relative differences  of gauge couplings then evolve purely 
logarithmically, exactly as in four dimensions.  Indeed, even the   
$\tilde b_i$-dependent logarithmic contributions in Eq.~(\ref{RGE}) 
are universal, and do not affect the unification of the couplings.
Thus, we find that when the $\tilde b_i$ are all equal,
we retain exactly the same logarithmic gauge coupling unification
that arises in four dimensions.

Despite this fact, it is still important to verify that the individual
gauge couplings themselves remain perturbative over the entire
energy range from $R^{-1}$ to $\Lambda\equiv M_{\rm GUT}$.  Otherwise, the use of
the one-loop result in Eq.~(\ref{RGE}) is no longer valid.
Towards this end, let us assume that $\tilde b_i\equiv \tilde b <0$ for
all $i$.
Since $\tilde b<0$, the power-law contributions to the gauge couplings 
push the couplings towards extremely weak values.  Indeed, in the limit 
where $\Lambda R\gg 1$, we find from Eq.~(\ref{RGE}) that
each of the gauge couplings scales in the ultraviolet as
\beq
   \alpha (\Lambda) ~\approx~  
        - {2\pi \delta\over \tilde b X_\delta} (\Lambda R)^{-\delta}~.
\label{asymcoupling}
\eeq
However, even though these couplings are extremely weak, the true loop
expansion parameter in such a situation is $\alpha_{\rm eff}\equiv N\alpha$
where $N\equiv X_\delta (\Lambda R)^\delta$ is the number of Kaluza-Klein
levels that have been crossed.  Indeed, $\alpha_{\rm eff}$ describes the
effective strength of the gauge interaction, since it characterizes the
coupling of each individual Kaluza-Klein mode multiplied by the multiplicity
of these modes.  
Thus, for true perturbativity, we must demand $\alpha_{\rm eff}\ll 4\pi$.

Remarkably, this constraint is satisfied no matter how large $\Lambda R$ becomes.
Indeed, we find that $\alpha_{\rm eff}\approx -2\pi \delta/\tilde b$ as $\Lambda R\to \infty$,
so that the condition for perturbativity becomes
                  $- \delta/ (2\tilde b) \ll 1$.
Thus, as long as $\tilde b$ is sufficiently large and negative, this condition
can be satisfied even if $\Lambda R \gg 1$.

As an example, let us consider  
a scenario in which, as discussed above, the zero-mode fields
are those of the MSSM and only
the GUT 
gauge bosons sit in the bulk.  For simplicity,
we shall take our unified gauge group to be $SU(5)$,
and
we shall also assume that $\delta=1$.
Since our low-energy theory is ${\cal N}=1$ supersymmetric,
the bulk fields necessarily fall into ${\cal N}=2$ supermultiplets.
Our bulk fields therefore consist of ${\cal N}=2$ vector multiplets transforming
in the adjoint
of $SU(5)$, leading to $\tilde b_i=\tilde b= -10$ for all $i$.
We then find that 
the effective gauge interaction strength at unification is 
$\alpha_{\rm eff}\approx 0.63$, which is considerably less than $4\pi$.
Note that this remains true even if $\Lambda R\approx 10^{13}$.  Thus
it is possible for the GUT
precursors to appear at the TeV scale even though the (logarithmic) gauge coupling 
unification does not occur until the usual scale $M_{\rm GUT}\approx 10^{16}$~GeV.

\begin{figure}[htb]
\centerline{
      \epsfxsize 3.4 truein \epsfbox {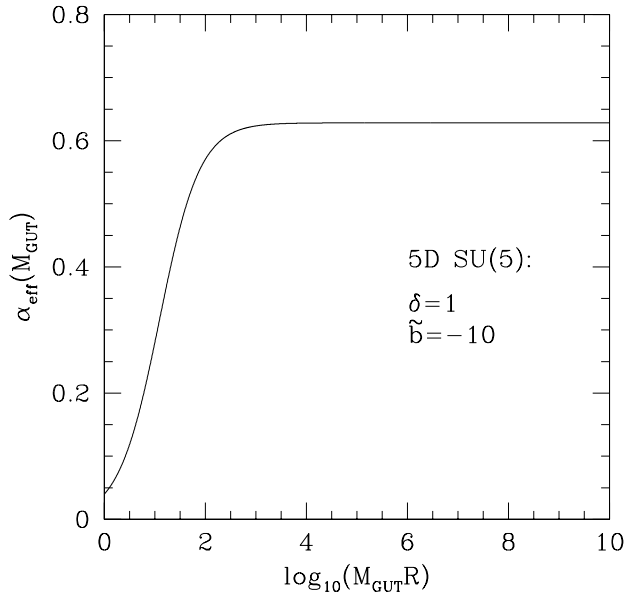}}
\caption{The effective unified coupling $\alpha_{\rm eff}(M_{\rm GUT})$ as
a function of $M_{\rm GUT}R$ for the five-dimensional $SU(5)$ GUT model
discussed in the text.  This coupling remains perturbative
for arbitrarily large values of $M_{\rm GUT}R$.}
\label{coupling}
\end{figure}
 
This behavior is illustrated in 
Fig.~\ref{coupling}, where we plot
the value of the effective
unified coupling $\alpha_{\rm eff}$ at $M_{\rm GUT}$
as a function of $R$, 
holding $M_{\rm GUT}$ fixed at its usual four-dimensional value
$2\times 10^{16}$~GeV.  We have taken $\delta=1$ and $\tilde b= -10$,
as discussed above.  It is clear that the effective coupling 
remains perturbative for arbitrarily large values of $M_{\rm GUT}R$,
saturating at its asymptotic value as early as $M_{\rm GUT} R\approx 100$.
Thus, the scale at which our GUT precursors appear can be 
separated by an arbitrary amount from the scale at which the
gauge couplings unify.

Note that this scenario requires
beta-function
coefficients $\tilde b_i$ which are
universal and {\it negative}\/.
Ordinarily, it might seem impossible to arrange $\tilde b_i$ negative
for each gauge group simultaneously, since in most scenarios (such as the
MSSM) the hypercharge coupling typically
becomes stronger with increasing energy.  However, $\tilde b_i$ is
negative in our scenario
precisely as a result of the GUT precursors.  
Indeed, in order to obtain a negative hypercharge beta-function coefficient, 
the $X$ and $Y$ gauge bosons are just what we require:  spacetime
vector bosons with non-trivial hypercharge assignments.  Thus, it is the
GUT precursors which permit the theory to remain perturbative
even as $\Lambda R\gg 1$.

One might worry that two-loop effects might be significant in such 
a scenario.
However, two-loop effects essentially {\it vanish}\/ in the
$\Lambda R\to \infty$ limit, since the presence of ${\cal N}=2$ supersymmetry
in the bulk ensures that the higher-loop power-law effects are suppressed by a factor
of $1/\Lambda R$ relative to the one-loop effects~\cite{DDG,KakuTaylor}.
Even when $\Lambda R$ remains finite, it is straightforward to verify that two- and
higher-loop corrections do not substantially alter the logarithmic
unification which emerges at one-loop order~\cite{DDG,KakuTaylor}.
Likewise, we remark in this context that the potentially damaging brane
surface kinetic terms discussed
in Ref.~\cite{Rattazzi} become vanishingly small in this context, since the
volume of the bulk becomes $10^{13}$ as large as the volume on the brane.
Thus, the unification of the gauge couplings is maintained.

It is also important to understand what happens if $\tilde b>0$.  In this
case, the gauge couplings become stronger rather than weaker as we evolve upwards
in energy, ultimately
hitting a Landau pole where $\alpha^{-1}(\Lambda)=0$.
We emphasize that this Landau pole is completely physical, since it
is induced by the cumulative effects of the Kaluza-Klein thresholds
as we evolve upwards in energy.
 [Even if we interpret Eq.~(\ref{RGE}) as a threshold correction
rather than as a renormalization group ``running,''
when $\tilde b >0$
this ``correction'' can push $\alpha^{-1}$ to negative values,
which is clearly unacceptable.]
In such cases, depending on the specific value of $\tilde b$, 
there is therefore a maximum allowed value of $M_{\rm GUT} R$
which can be tolerated
if we imagine varying $R$ while holding $M_{\rm GUT}$ fixed.
For example, in some of the ``minimal'' models of Ref.~\cite{HN}, the bulk contains
not only the $SU(5)$ gauge bosons, but also two ${\bf 5}$ representations
(for the Higgses) and four ${\bf 10}$ representations (for the first two
generations).  Since these are all ${\cal N}=2$ supermultiplets,
this leads to a value $\tilde b = +4$.  
(In general, with $n_5$ multiplets transforming in the {\bf 5} representation
and $n_{10}$ multiplets in the {\bf 10}
representation, one obtains $\tilde b= -10+n_5 + 3n_{10}$.)
We then find that perturbativity
in such models requires a maximum value $M_{\rm GUT} R \lsim 44$ 
in order to avoid a Landau pole (or equivalently, to
avoid negative inverse gauge couplings).  
Since some of the models in Ref.~\cite{HN}
require $M_{\rm GUT} R\approx 100$, 
it is unclear how the perturbative logarithmic unification 
prediction is maintained in such models.

\section{Power-Law Running and\\ Non-Trivial Ultraviolet Fixed Points}

In the setup described in the previous section, 
the asymptotic ultraviolet power-law scaling of gauge couplings   
towards weak values is exactly compensated by the asymptotic power-law growth of the
number of degrees of freedom in the theory   
in such a way that the product of these two quantities remains a constant.
Thus, the effective strength of the gauge interactions appears to
approach a non-trivial fixed point in the ultraviolet.
Such behavior for gauge couplings with $\tilde b <0$ 
was also observed previously in Ref.~\cite{Agashe}.
Of course, in our particular scenario we do not need to approach 
the truly asymptotic limit
$\Lambda R\to \infty$ because the logarithmic terms in Eq.~(\ref{RGE})
eventually induce a unification at $\Lambda\equiv M_{\rm GUT}$;
we therefore can take this as a cutoff for the higher-dimensional
running.
Nevertheless, it is intriguing to consider the formal limit in which
we disregard issues of unification and take $\Lambda R\to \infty$.  

It is already apparent from Fig.~\ref{coupling}
that as long as $\Lambda R \gsim 100$,
our theory essentially becomes ``scale invariant'' in the 
sense that the ultraviolet physics becomes independent of the low-energy
scale $R^{-1}$ at which the GUT precursors appear.
We may also rephrase this observation
directly in terms of the effective couplings $\alpha_{{\rm eff},i}\equiv N \alpha_i$
where $N\equiv X_\delta (\Lambda R)^\delta$.
Given the evolution equations for the couplings $\alpha_i$ in
Eq.~(\ref{RGE}), it is straightforward to show that 
the effective couplings $\alpha_{{\rm eff},i}$ evolve 
according to
\beqn
  \Lambda {d \alpha^{-1}_{{\rm eff},i} \over d\Lambda} &=& 
    - \left( \delta \alpha^{-1}_{{\rm eff},i} + {\tilde b_i \over 2\pi}\right)
    + \left( {\tilde b_i - b_i \over 2\pi X_\delta}\right) (\Lambda R)^{-\delta}\nonumber\\
    &&  ~~~ + {c_i\over 2\pi}\, {\alpha_{{\rm eff},i}\over 4\pi}\, (\Lambda R)^{-\delta}
     + ...~, 
\label{RGEeff}
\eeqn
where in the second line we have written the dominant
two-loop contributions arising from the bulk and boundary fields
running in the loops (with $c_i$ representing a two-loop beta-function coefficient). 
Thus, even though the individual gauge couplings $\alpha_i$ themselves
evolve with power-law behavior, 
we see from Eq.~(\ref{RGEeff}) that for $\Lambda R\gg 1$, 
the {\it effective}\/ gauge couplings
$\alpha_{{\rm eff},i}$ each approach an ultraviolet fixed point at
$\alpha_{{\rm eff},i} = -2\pi\delta / \tilde b_i$.
Moreover, if $\tilde b_i \equiv \tilde b$ for all $i$,
we see that even though the differences of the  
gauge couplings continue to evolve logarithmically,
the fixed-point values of the {\it effective}\/ gauge couplings  
all become equal.  
Thus, in this sense, we see that the effective strengths of the
gauge interactions in this theory each 
flow to a {\it common}\/ fixed point in the ultraviolet.
Note that two- and higher-loop effects merely contribute
additional power-law terms in Eq.~(\ref{RGEeff}) which again vanish
in the $\Lambda R\to \infty$ limit.  Such contributions therefore
do not alter the ultraviolet fixed-point structure of these theories.

It is natural to interpret these results as indicating
the emergence of a non-trivial (interacting)
ultraviolet fixed point corresponding 
to a supersymmetric, higher-dimensional, unified gauge theory.
Indeed, such higher-dimensional fixed-point gauge theories
are known to exist in uncompactified 
five~\cite{Seiberg5} and six~\cite{Seiberg6} dimensions.
Since we expect the ultraviolet (short-distance) limit 
of our compactified theory to reproduce the physics of an 
uncompactified higher-dimensional theory, it is tempting to identify the  
ultraviolet limit of our theory as one
of the interacting fixed-point theories discussed in
Refs.~\cite{Seiberg5,Seiberg6}.

As an example, let us consider the case of $SU(N)$ gauge theory
in five dimensions.  If the only matter consists 
of $n_f$ ``quarks'' transforming in the fundamental representation,
then the necessary and sufficient condition~\cite{Seiberg5} for the 
existence of an interacting ultraviolet fixed point is $n_f\leq 2N$.
This is equivalent to our requirement that $\tilde b\leq 0$.

It is important to stress that for unitary groups, the conditions 
for the existence of
such non-trivial fixed points are generally {\it stronger}\/ than 
merely demanding $\tilde b<0$.  
For example, if we also introduce 
bulk matter which transforms in the {\it antisymmetric tensor}\/ 
representation [\eg, the {\bf 10}~representation
of $SU(5)$], then further restrictions on the number of fundamental
representations
must be imposed in order to 
guarantee the fixed-point behavior of the theory in the ultraviolet~\cite{Seiberg5}.
These conditions are consistent with $\tilde b\leq 0$, but
provide further constraints on the precise matter content.
However, our main point is that 
we can always ensure that our
theory flows to a non-trivial fixed point in the ultraviolet
by choosing the bulk field content appropriately. 
This then guarantees the self-consistency of our scenario and
the large separation in energy scales that it contains. 

One important by-product of this analysis is that it essentially
furnishes us with an alternative, four-dimensional ``Kaluza-Klein'' realization 
of these fixed-point theories.
In such a realization, 
the effective higher-dimensional 
gauge coupling at the fixed point 
asymptotically emerges in the ultraviolet
as the product $\alpha_{\rm eff}= N\alpha$. 
[Note that the {\it dimensionful}\/ gauge coupling
in higher dimensions is 
$\alpha_{4+\delta}(\Lambda) = 
 \Lambda^{-\delta} \alpha_{\rm eff}(\Lambda)$.] 
Moreover,
the existence of such ``Kaluza-Klein'' realizations for these
higher-dimensional fixed points
may provide a new tool for studying
the properties of these
fixed points under compactification, orbifold projection,
GUT symmetry breaking, and even supersymmetry breaking.
These are precisely the issues that must be addressed
if such higher-dimensional fixed-point
gauge theories are to have
phenomenologically relevant compactifications to four
dimensions. 
Thus, these four-dimensional ``Kaluza-Klein'' realizations
of these interacting fixed points may provide a new approach which 
can be used when studying how these theories behave under
phenomenologically relevant compactifications,
and when calculating the subleading corrections 
that such compactifications introduce.

\section{Deconstruction and UV/IR Fixed-Point Matching}

Recently, a new four-dimensional  ultraviolet completion of 
Kaluza-Klein theories was proposed~\cite{acg}.
When generalized to a supersymmetric context~\cite{cegk},
this ``deconstruction'' proposal embeds the Kaluza-Klein theory
into a four-dimensional asymptotically free
theory based on the gauge group $SU(N_c)^{N+1}$.
The model can also be described by a ``moose'' (or quiver) diagram with $N+1$ sites
and links.
At each site on the moose diagram, there is an ${\cal N}=1$ supersymmetric $SU(N_c)$ gauge
theory;
likewise, the links represent  
chiral multiplets $\Phi_I$ which transform in the bifundamental representation
of the two adjacent $SU(N_c)$ gauge groups.
Each site therefore contains $N_f=N_c$ flavors transforming 
in the fundamental representation of the $SU(N_c)_i$ gauge group.
At the deconstruction scale, the scalar components $\phi_I$ within the chiral 
supermultiplets $\Phi_I$ each 
accrue equal vacuum expectation values $\langle\phi_I\rangle=v$
which in turn break the
gauge group to the diagonal:  $SU(N_c)^{N+1} \rightarrow SU(N_c)$.
In the energy range $gv/N < \mu < gv$, the resulting spectrum is 
then similar to that of a pure ${\cal N}=2$ supersymmetric 
$SU(N_c)$ Yang-Mills theory
compactified on a circle.

Deconstruction naturally provides a renormalizable
context for analyzing the ultraviolet fixed points discussed in this paper. 
Note, first of all,
that the effective coupling $\alpha_{\rm eff}$ which approaches the ultraviolet fixed-point value
represents in the deconstruction scheme the gauge coupling of gauge groups at each site
at the deconstruction scale. 
Moreover, the fact that $\alpha_{\rm eff}$ is independent of $N$
is natural from the deconstruction viewpoint. 

However, when embedding these ultraviolet fixed points into a deconstructed
theory at even higher energy scales, 
it is important to maintain the ultraviolet insensitivity
of low-energy observables.
Otherwise, the underlying fixed-point structure will be destroyed by the deconstruction. 
Thus, in order to preserve this ultraviolet insensitivity,
we see that the ultraviolet fixed points from the Kaluza-Klein perspective
must somehow be identified with (or matched with) 
 {\it infrared}\/ fixed points of the theory above the deconstruction scale.

The obvious candidates for these infrared fixed points are the four-dimensional
fixed points proposed in connection with the conjectured ${\cal N}=1$
Seiberg duality~\cite{seiberg}. For $SU(N_c)$ gauge groups, it was argued in Ref.~\cite{seiberg}
that infrared fixed points exist
only when the number $N_f$ of flavors satisfies $ 3N_c/2 < N_f < 3N_c$.
Thus, in building our deconstructed theory, 
we are led to introduce  an additional $n_f > N_c/2$ flavors
of chiral multiplets 
$Q_i, {\tilde Q}_i$ at each site.
The total number of flavors at each site is
therefore $N_f=N_c+n_f$. 
For our purposes we will not need a full superconformal
field theory in the infrared limit of our deconstructed theory, 
and it suffices merely to impose the existence of an infrared
fixed point for the gauge couplings. 
In our case, the exact beta-function becomes~\cite{SV} 
\beq
        \beta(g) ~=~  {g^3\over 16\pi^2}\, 
       {b' \over  1- N_c g^2 /8\pi^2}
\label{k1}
\eeq
where the coefficient $b'$ is given by  
\beq
        b' = -2N_c+n_f - {N_c \over 2} (\gamma_{\Phi}^{I-1}+\gamma_{\Phi}^{I}) -
         {n_f \over 2} (\gamma_{Q}^{I}+\gamma_{\tilde Q}^{I}) 
\label{k2}
\eeq
and where the one-loop anomalous dimensions 
of the fields at each site/link $I$ are given by
\beqn
    \gamma_{\Phi} &=&
         -{1 \over 8 \pi^2} \left(2 {N_c^2-1 \over N_c}g^2 -n_f \lambda^2\right)~, \nonumber\\
     \gamma_{Q} = \gamma_{\tilde Q}  &=& -{1 \over 8 \pi^2} \left({N_c^2-1 \over N_c} g^2
     - N_c \lambda^2\right)~.
\label{k3}
\eeqn
The superpotential describing
the deconstructed 
theory is given by
\beq
     W = \sum_{i, I} \, (\sqrt{2}\lambda {\tilde Q}_{i,I} \Phi_I Q_{i,I+1}- m {\tilde Q}_{i,I} Q_{i,I})~,
\label{d2}
\eeq
whereupon we see that the Kaluza-Klein spectrum for the $n_f$ hypermultiplets is recovered 
by setting $m= \sqrt{2} g^2 v/\lambda$. 
In addition, the zero mode of the hypermultiplets has a mass 
$m_0^2 = 2 v^2 (g^2-\lambda^2)^2/\lambda^2$. 
Since no site is singled out, we have taken
$\gamma_{\Phi}^{I} \equiv \gamma_{\Phi}$ for all $I$
in Eq.~(\ref{k3}).
We then find that
requiring $\beta(g)=0$ 
defines a curve of fixed points
in the $(g,\lambda)$ plane.

Let us now consider how to match
the ultraviolet fixed point of the five-dimensional
theory with the infrared fixed point of the deconstructed theory.
First, we must take $\lambda=g$ in our deconstructed
theory in order to guarantee that the interactions below the deconstruction
scale respect ${\cal N}=2$ supersymmetry.
Second, since our theory below the deconstruction scale has
$SU(N_c)$ gauge symmetry with $n_f$ flavors,  we see that $\tilde b= -2N_c+n_f$.
Thus, our Kaluza-Klein theory has $g^2_{\rm eff}= 8\pi^2 /(2N_c-n_f)$.

In order to match the ultraviolet and infrared fixed points, 
we now require solutions for $\beta(g)=0$ 
to occur at $g=g_{\rm eff}$.
Setting $b'=0$ and requiring $g=g_{\rm eff}$, we obtain the constraint
\beq
     (2 N_c - n_f ) ( n_f - N_c) ~=~  2 + {n_f\over N_c}~.
\label{d4}
\eeq
In the large-$N_c$ limit, this has two solutions:  $n_f=N_c$ and $n_f= 2N_c$.
We can reject the second solution because it implies $\tilde b=0$, or 
$g_{\rm eff}\to \infty$.  (Equivalently, taking $n_f = 2N_c$ violates the
infrared fixed-point constraint $N_f < 3N_c$ in the original deconstructed theory.)  
However, the solution with $n_f=N_c$ yields $N_c \alpha_{\rm eff} = 2\pi$,
which is finite and semi-perturbative.
Of course, we must be slightly more careful because this value for 
$g_{\rm eff}$ causes the denominator in Eq.~(\ref{k1}) to diverge.
However, for any large but finite value of $N_c$, there exists an exact solution 
for $n_f/N_c$ in Eq.~(\ref{d4}) which is only slightly larger than $1$.
For this exact solution, we find that $b'=0$ and consequently $\beta(g)=0$.
Note that higher-loop contributions in Eq.~(\ref{k3}) are suppressed by 
the small gauge coupling and thus provide only a small correction to the 
exact solution for $n_f/N_c$ in Eq.~(\ref{d4}). 

Of course, a fully non-perturbative 
solution for the ultraviolet/infrared fixed-point matching
might also exist.
In particular, if we demand that the infrared limit of
our deconstructed theory is actually superconformal, 
then it may be possible
to achieve non-perturbative solutions to the constraint $b'=0$
if the full, non-perturbative, anomalous dimensions satisfy  
\beq
       \gamma_\Phi = -2~,~~~~~~~ \gamma_Q + \gamma_{\tilde Q} = 2~.
\label{nonpert}
\eeq 
The non-perturbative solution to these equations should therefore 
match the ultraviolet fixed point of our Kaluza-Klein theory in order
to preserve the ultraviolet insensitivity of the low-energy theory.

Thus, we conclude that it is possible, in principle,
to deconstruct our ultraviolet fixed
points while maintaining their ultraviolet insensitivity.
Note that the embedding into a deconstructed theory also ensures that there are no
other non-renormalizable operators which could appear 
in the ultraviolet theory
and cause difficulties when extrapolated down to lower energies.

\section{Orbifold Choices and Non-Universal Logarithms} 

Thus far, we have presented a general GUT scenario in which GUT precursor
states can be extremely light compared with the scale of gauge coupling unification.
Our purpose has been to illustrate the emergence and utilization of non-trivial fixed points 
as a means of incorporating widely separated energy 
scales within a single model.  
However, in order to build a fully consistent model, we must choose an explicit
orbifold and take into account certain additional contributions that arise~\cite{HN,corresp,talk,zurab}.
    
Let us first consider the case of the five-dimensional $SU(5)$ 
theory compactified
on the $S_1/\IZ_2$ orbifold presented in Sect.~II.
Under the $\IZ_2$ action, Standard-Model gauge supermultiplets are even while  
GUT supermultiplets such as the $X$ and $Y$ gauge supermultiplets
are odd.  However, since this is a five-dimensional theory,
the $X$ and $Y$ supermultiplets also have fifth components $X_5$ and $Y_5$, and
consistency of the $\IZ_2$ orbifold requires that these components be even.
The zero modes of such fields then produce
additional non-universal logarithmic contributions to the runnings of the
gauge couplings, with coefficients $(b_1,b_2,b_3)=(5,3,2)$.
However, it is easy to verify that taking $R^{-1}\sim$~TeV still leads to an 
approximate unification at $M_{\rm GUT}\approx
2\times 10^{13}$~GeV which, although not as precise as the MSSM unification,
is nevertheless more precise than the unification in the Standard Model.
Remarkably, as we shall discuss in Sect.~VII, such an extra dimension also lowers
the fundamental Planck scale to this new unification scale.
Of course, at a phenomenological level,
it still remains necessary to find a mechanism to give 
masses to these $X_5$ and $Y_5$
zero modes, 
since they represent colored and fractionally charged scalars.

Another option is to compactify our theory on an $S_1/(\IZ_2\times \IZ_2')$
orbifold with two distinct $\IZ_2$ discrete actions~\cite{Kawamura,HN} associated
with $y\to -y$ and $y\to \pi R-y$. 
Under the $(\IZ_2,\IZ_2')$ orbifold actions, the Standard-Model gauge fields $A_\mu$
have $(+,+)$ eigenvalues [with corresponding cosine modings as
in Eq.~(\ref{KKdecomp}) with $n\in 2\IZ$ only],
while $A_5$ has $(-,-)$ eigenvalues (resulting in sine modings with $n\in 2\IZ$). 
Likewise, the GUT precursors $X_\mu$  
and $Y_\mu$ have $(-,+)$ eigenvalues (resulting in sine modings
with $n\in 2\IZ+1$),  
while $X_5$ and $Y_5$ have $(+,-)$ eigenvalues (resulting
in  cosine modings with $n\in 2\IZ+1$).
With this orbifold choice, only the Standard Model fields
have zero modes, but this occurs at the expense of 
splitting the complete GUT multiplets 
at each Kaluza-Klein level into a subset at even levels, 
with beta-function
coefficients $(\tilde b_1,\tilde b_2,\tilde b_3)=(0,-4,-6)$,
and a subset at odd levels, with 
$(\tilde b'_1,\tilde b'_2,\tilde b'_3)=(-10,-6,-4)$.
This results in a staggered, ``zig-zag'' running 
for the gauge couplings
which averages to a universal power-law running 
with an effective radius $R/2$,
along with a non-universal logarithmic correction.
Specifically, the gauge couplings now run according to  
\beqn
       \alpha_i^{-1} (\Lambda)  &\approx &
       \alpha_i^{-1} (M_Z) 
      - {b_i\over 2\pi} \ln {\Lambda\over M_Z} +
     {\tilde b\over 4\pi} \ln {{\Lambda R}\over 2} \nonumber\\
        && ~~~~ -  {\tilde b\over 2\pi} \left\lbrack
          \left({\Lambda R\over 2}\right) -1 \right\rbrack  
      - {\tilde b'_i \over 2\pi } Y
\label{newRGE} 
\eeqn
where $\tilde b\equiv \tilde b_i + \tilde b'_i= -10$ for all $i$, 
where we have neglected various universal additive constants, 
and where the non-universal logarithm is given by 
\beq
        Y ~\equiv~ \sum_{n=0}^{(\Lambda R-2)/2} \ln{ 2n+2\over 2n+1}
     ~\approx~ \half \ln { \pi\Lambda R\over 2}~
\eeq 
with the last approximation holding in the $\Lambda R\gg 1$ limit.
Given this running, we then find that the three gauge couplings
continue to experience an approximate unification.
With $R^{-1}\sim$~TeV, the unification scale
is unfortunately quite high ($M_{\rm GUT}\approx 10^{21}$~GeV),
but increasing $R^{-1}$ not only improves the accuracy of the resulting unification
but also lowers the unification scale.  Asymptotically, with $R^{-1}\approx 10^{15}$~GeV,
we obtain an essentially {\it exact}\/ unification at $M_{\rm GUT}\approx 10^{17}$~GeV.
Although this resembles the energy scales in the scenario 
in Ref.~\cite{HN}, we stress that
the bulk theory here contains only gauge fields, 
and our gauge couplings become {\it weak}\/ rather 
than strong in the ultraviolet limit.  
Of course, one can also place additional matter in the bulk or
on the brane so as to increase the accuracy of the unification
and lower the unification scale~\cite{zurab}.

A final option is to place a single Higgs five-plet in the bulk.
After compactifying on the $S_1/(\IZ_2\times \IZ_2')$ orbifold, 
the bulk Higgs doublets (triplets) are at even (odd) Kaluza-Klein levels.
We thus have 
$\tilde b= -9$,
$(\tilde b_1,\tilde b_2,\tilde b_3)=(3/5, -3,-6)$,
and $(\tilde b'_1,\tilde b'_2,\tilde b'_3)=(-48/5,-6,-3)$.
Note that these values of $\tilde b_i$ are the same as those
of Ref.~\cite{DDG}. 
This yields a unification
of gauge couplings which is similar to (and approximately as accurate as) the case
with only gauge bulk fields discussed above;
in each case, small (few-percent) threshold corrections at the 
unification scale are sufficient to render the unification exact
for all values of $R^{-1}$.
Of course, in each case one must place the remaining Higgs field(s) 
on a Standard-Model brane lacking the GUT symmetry in order to 
avoid the doublet/triplet problem.

\section{Discussion}
  
In this paper, we have shown that it is possible for certain ``GUT precursor''
states to appear with masses that are significantly below the
scale at which grand unification occurs.  
As we have seen, the fixed-point structure 
which makes this possible
can be viewed
as a general tool which permits the construction
of generalized ``hybrid'' models in which widely separated energy
scales can coexist in a natural way. 
  
Needless to say, these observations prompt a number of important questions,
both phenomenological and theoretical.
Among the most important phenomenological questions is the issue of proton
decay.  Ordinarily, light $X$ and $Y$ gauge boson precursors will mediate
rapid proton decay.  
However, as in all low-scale extensions to the Standard Model, this 
problem may be cured through the use of split fermions on the branes~\cite{split}
or through the introduction of extra discrete 
symmetries~\cite{DDG,Kakushadze}.
Likewise, other phenomenological issues include doublet/triplet splitting
and general issues of flavor physics.  Although we have not attempted
to make a complete GUT model that accommodates these phenomena, one could
imagine doing so following the lines of Refs.~\cite{Kawamura,HN,JMR,otherHN}  
except that we now have the interesting option of extending the energy scales of such models 
into the TeV range.  Moreover, we are also free to introduce further matter
into the bulk beyond what we have discussed here, 
provided we ensure that $\tilde b$ remains sufficiently large
and negative and provided the conditions~\cite{Seiberg5} for an ultraviolet
fixed point are maintained.
Indeed, the presence of two widely separated scales in such models
suggests that we might even try to  
use the GUT precursors to trigger electroweak symmetry breaking
and/or supersymmetry breaking.
Of course, we have not speculated on the unknown 
dynamics which might ultimately be responsible for generating and
stabilizing such a large radius;  ideas along these lines can be
found, \eg, in Ref.~\cite{stabilize}.

Our results in this paper also raise a number of theoretical issues. 
The most important concerns the ultraviolet limit of our theory.
Although we have shown that the evolution of the gauge couplings is consistent
with perturbativity even when the effective higher-dimensional energy interval 
is large, one must actually verify that {\it all}\/ correlation functions
in the theory remain finite and under control over this large energy range.
This is clearly connected with the over-riding question discussed in
Sect.~IV concerning the manner in which we approach a scale-invariant fixed 
point in the ultraviolet.   
By counting Kaluza-Klein states and vertex factors in diagrams with arbitrary numbers
of loops and external legs,
it is straightforward to demonstrate that 
all diagrams in this theory
necessarily scale as $(N \alpha)^k \sqrt{\alpha}^\ell$ where 
$k$ and $\ell$ are non-negative integers.
Thus, in the ultraviolet limit, such diagrams either vanish (if $\ell\not= 0$)
or approach a fixed finite value (if $\ell=0$).
Indeed, even though the number of states in this theory is diverging
at higher energies, the individual gauge couplings are falling to zero
in an exactly compensatory manner.

For example, the four-fermion amplitude 
for tree-level Kaluza-Klein exchange in the $\delta=1$ case
becomes
\beqn
    A(s) ~&\sim&~ \alpha(s) \, \sum_n {1\over s-n^2/R^2} 
    ~\sim~     {\alpha(s)\over \sqrt{s}} \,R \cot(\pi R\sqrt{s}) \, 
      \nonumber\\
        &&  ~\longrightarrow~ {\alpha_{\rm eff} \over s} \,\cot(\pi R\sqrt{s}) 
\label{amplitude}
\eeqn
where we have taken the limit $s R^2 \gg 1$ and identified
$\alpha_{\rm eff} \sim (R\sqrt{s}) \alpha(s)$ as $s\to \infty$.
Thus, since $A(s)$ continues to have 
the asymptotic energy dependence $\sim 1/s$,
no unitarity bounds are violated in the ultraviolet.
(The divergence when $R\sqrt{s}\in\IZ$
merely reflects a fine-tuned Kaluza-Klein resonant pole.)
Moreover, as we have already noted,
the presence of ${\cal N}=2$ supersymmetry in the bulk
ensures the vanishing of many diagrams which would otherwise lead
to large and potentially uncontrollable corrections. 
It will therefore be interesting to explore the ultraviolet self-consistency 
of this theory further~\cite{poppitz}.
We stress, however, that amplitudes such as those in Eq.~(\ref{amplitude})
will remain consistent with unitarity bounds only as long as the 
growing number of Kaluza-Klein states is offset by the power-law running 
of the gauge couplings.
If there are more than two extra dimensions, the power-law running for the
gauge couplings will be cancelled by the ${\cal N}=4$ supersymmetry
that would be required in the bulk.  In such cases, the theory will
remain unitary only up to the scale at which these extra dimensions
become apparent.

Note that the presence of a large, scale-invariant energy interval 
in our model is reminiscent of previous models~\cite{FramptonVafa}
in which ``conformality'' is invoked to solve the technical hierarchy
problem.  These models also share certain features with GUT orbifold models    
formulated in warped backgrounds~\cite{Pomarol}, as well as with
string models in which physics at all scales conspires 
to eliminate the quantum-mechanical sensitivity between light and  heavy 
energy scales~\cite{missusy}.

Another important theoretical issue for our models concerns gravity.
In this paper we have merely proposed a model of gauge interactions, and as such,
any value of $R$ for the GUT precursors is permitted.  However, if we wish 
to incorporate gravity, further complications arise.
Since our large extra dimension is presumably also felt by gravity, 
Kaluza-Klein gravitons will induce Newton's constant to run more quickly.
The standard Gauss-law arguments of Ref.~\cite{TeV2} then imply
that taking $R^{-1}\sim {\cal O}$(TeV) 
lowers the effective higher-dimensional Planck scale 
$M_\ast$ to approximately
$10^{14}$~GeV.
Although this is comfortably within all experimental constraints,
this value is slightly below $M_{\rm GUT}\approx 10^{16}$~GeV.
This indicates that we reach a region of strong gravity {\it before}\/
our gauge couplings unify.  

There are two ways in which this situation might be avoided.
Within the context of the $S_1/\IZ_2$ orbifold discussed in 
Sect.~VI, we have seen that the additional non-universal
logarithmic contributions actually lower the unification scale
to approximately $10^{13}$~GeV.  
Thus, for the $S_1/\IZ_2$ orbifold, there is no difficulty.
On the other hand,
for the $S_1/(\IZ_2\times \IZ'_2)$ orbifold, 
we have seen in Sect.~VI that the unification
scale generally exceeds $10^{16}$~GeV.
One possible solution is then to restrict the GUT
precursor scale to the range $R^{-1}\gsim 10^{10}$~GeV;  this 
ensures that $M_\ast \gsim M_{\rm GUT}$.
Such GUT precursors would still be significantly lighter than the
unification scale, although no longer accessible to upcoming collider experiments.
Another solution, however,
is to lower the value of $M_{\rm GUT}$ to 
$10^{14}$~GeV
by introducing further states with appropriate gauge quantum numbers
into the theory.  Such states would not spoil the universal power-law
behavior of this theory if they are restricted to the orbifold fixed points.
Gauge coupling unification in such a scenario would then continue to be
logarithmic, as desired.

A similar issue arises if we attempt to embed this scenario
into string theory.  Let us first consider the case of the 
perturbative heterotic string.
In general, the heterotic string scale is related to the Planck scale
through the unified gauge coupling:
$M_{\rm string}= g_{\rm GUT} M_{\rm Planck}$.
However, as a result of the asymptotic power-law running of the gauge couplings,
we know that $\alpha_{\rm GUT}^{-1} \sim (M_{\rm string} R)^\delta$.
(In writing this relation, we are imagining an ultimate identification
of $M_{\rm string}$ with $M_{\rm GUT}$.)
However, combining these results and neglecting all numerical factors, 
we find that $M_{\rm string} \sim M_\ast$ where 
$M_\ast^{2+\delta} \sim M_{\rm Planck}^2/R^\delta$ is the 
higher-dimensional Planck scale at which gravitational effects
become significant.
Thus, for the heterotic string, the power-law scaling of the gauge coupling
implies that it is not possible to separate
the scales at which both gravitational and string-theoretic effects become
significant.  Note that this result persists regardless of the value
of the underlying (perturbative) string coupling. 

Within the context of Type~I strings, however, 
this situation is somewhat different.
For Type~I strings, the relation between the string and Planck scales is
modified:  $M_{\rm string} = g_{\rm GUT} \sqrt{g_{\rm string}} M_{\rm Planck}$,
where $g_{\rm string}$ is the underlying string coupling.  
Again taking $\alpha_{\rm GUT}^{-1} \sim (M_{\rm string} R)^\delta$,
we now find the relation
\beq
      M_{\rm string}^{2+\delta} ~\sim~ g_{\rm string} \,M_\ast^{2+\delta}~.
\label{TypeIrelation}
\eeq
Thus, since $M_\ast\lsim 10^{14}$~GeV (with the maximum value for $M_\ast$ occurring
for $\delta=1$), we see that we can indeed take $M_{\rm string}\approx M_{\rm GUT}$
if we have a large string coupling $g_{\rm string}\gg 1$.
This suggests that embedding our scenario within the context of Type~I strings
necessarily involves strong couplings and non-perturbative
physics.  
Moreover, since the Type~I string coupling is 
given by $g_{\rm string}\sim \alpha_{\rm GUT} V_6 M_{\rm string}^6$
where $V_6$ is the {\it total}\/ six-dimensional compactification volume,
we see that taking $g_{\rm string}\gg 1$ requires
$M_{\rm string}^6 V_6 \gg M_{\rm string}^\delta V_\delta$
where $V_\delta$ is that portion of the compactification volume  
which produces power-law running for $\alpha_{\rm GUT}$.

\medskip

Thus, to summarize, we have shown that it is possible for ``GUT precursor''
states to appear with masses that are extremely light compared with the scale
of gauge coupling unification.
This suggests a possible new TeV-scale direction 
for orbifold GUT models. 
Indeed, more generally,
we have seen that ultraviolet embeddings into fixed-point theories
can be used to provide a new method for maintaining or stabilizing
a wide separation of energy scales within a single model.
Using this technique, ``hybrid'' models with coexisting high and low energy scales
can therefore be constructed in a variety of contexts.
Equally importantly, however, our four-dimensional ``Kaluza-Klein'' realization
of such fixed points should also provide a new technique for the study
of such theories and their properties under various compactifications,
both with and without supersymmetry breaking and gauge symmetry breaking.
These and other directions await exploration.

\section*{Acknowledgments}

KRD was supported in part by the National Science Foundation
under Grant PHY-0071054, and by a Research Innovation Award from 
Research Corporation.  
ED was supported in part by the RTN European Program
HPRN-CT-2000-00148.
We wish to thank C.~Grojean, K.~Intriligator, G.D.~Kribs, 
J.~Mourad, M.J.~Strassler 
and especially E.~Poppitz for discussions. 
KRD and TG wish to acknowledge the hospitality of
the Aspen Center for Physics where this work was initiated,
and KRD and ED wish to acknowledge the hospitality of the Theoretical Physics
Institute at the University of Minnesota where this work was completed.


\vskip -0.2 truein



\end{document}